

\magnification=1200
\parindent 1.0truecm
\baselineskip=16pt
\rm
\null


\footline={\hfil}
\vglue 0.8truecm
\rightline{\bf DFUPG--68--95 }
\rightline{\sl April 1995 }
\vglue 2.0truecm

\centerline{\bf Low--Energy Kaon--Nucleon Interactions and Scattering }
\centerline{\bf at DA$\Phi$NE$^{\dag}$ }

\vglue 1.0truecm
\centerline{\sl Paolo M. Gensini }
\centerline{\sl Dip. di Fisica dell\'\ Universit\`a di Perugia, Perugia,
Italy, and }
\centerline{\sl I.N.F.N., Sezione di Perugia, Italy }

\vglue 6.0truecm
\centerline{ A chapter to be included in }
\centerline{\sl The Second DA$\mit\Phi$NE Physics Handbook }
\centerline{ (by the DA$\Phi$NE Theory Working Group) }
\centerline{ ed. by L. Maiani, G. Pancheri and N. Paver (INFN, Frascati 1995).}
\vglue 3.0truecm
\hrule
\vglue 0.5truecm
\item{\dag)} Research supported by the E.E.C. Human Capital and Mobility
Program under contract No. CHRX--CT92--0026.
\pageno=0
\vfill
\eject



\footline={\hss\tenrm\folio\hss}

\vglue 0.5truecm
\centerline{\bf Low--Energy Kaon--Nucleon Interactions and Scattering }
\centerline{\bf at DA$\Phi$NE$^{\dag}$ }

\vglue 1.0truecm
\centerline{\sl Paolo M. Gensini }
\centerline{\sl Dip. di Fisica dell\'\ Universit\`a di Perugia, Perugia,
Italy, and }
\centerline{\sl I.N.F.N., Sezione di Perugia, Italy }

\vglue 3.0truecm

\leftline{\bf Abstract. }
\vglue 0.3truecm

We present in this contribution the basic formul\ae\ for the analysis of
low--momentum charged-- and neutral--kaon interactions in hydrogen, including
as well a (brief) description of the problems left open by past experiments,
and of the improvements DA$\Phi$NE can be expected to offer over them.
Interactions in deuterium and other light nuclei will be only briefly
mentioned, and only in those respects touching directly upon the more
``elementary'' aspects of kaon--nucleon interactions.

\vglue 1.0truecm
\leftline{\bf 1. Introduction. }
\vglue 0.3truecm

DA$\Phi$NE is expected to produce, with ``standard'' assumptions about
luminosity and cross sections, about $1.25 \times 10^{10}\ K^\pm$'s and
$8.5 \times 10^9\ K_L^0$'s per year of operation, considering a conventional
``Snowmass year'' of $10^7\ s$. With a detector of KLOE's size one can thus
expect to observe up to millions of interactions per year,
even in the low--density gas (mostly helium at atmospheric pressure)
filling its fiducial volume.

A first question has thus to be answered: do these events contain useful
physics to be worth recording and interpreting? One could even go further
and envisage using DA$\Phi$NE as a source of high--resolution ($\Delta p/p$
of respectively $1.1 \times 10^{-2}$ for $K^\pm$'s and $1.5 \times 10^{-2}$
for $K_L^0$'s), low--momentum kaons ($127\ MeV/c$ for the former, minus the
losses in the matter along the kaon path from DA$\Phi$NE's beam--beam
interaction region to the point where it interacts in the detector, and
$110\ MeV/c$ for the latter), to measure with a {\sl dedicated} detector
$KN$ and $\overline{K}N$ interactions, e.g. in a toroidal volume
filled with gaseous $H_2$ or $D_2$, and possibly also with heavier--element,
gaseous (i.e. low--density) targets.

The machine has thus the ability to explore a small kinematical region
($90\ MeV/c \leq p_{lab} \leq 120\ MeV/c$), very little investigated in
the past:
only few bubble--chamber $K^\pm$ experiments in hydrogen (and
deuterium)$^{1,2}$ plus very few data points on $K_S^0$
regeneration$^3$ exist,
all with extremely low statistics and more than a decade old (the last
experiment$^2$ to cover this region was carried out in the second half
of the seventies by the TST Collaboration at the hydrogen bubble chamber
in NIMROD's low--energy kaon beam). Since in dispersive calculations
of low--energy parameters for $KN$ interactions ($\overline{K}N Y$
and $\pi Y Y'$
coupling constants, scattering lengths, $\sigma$--terms) the bulk of the
uncertainties comes from the integrals over the unphysical regions, for
whose description one must extrapolate data analyses for quite an energy
range, down from the physical region well above the
charge--exchange threshold, DA$\Phi$NE, as the cleanest, lowest--energy kaon
{\sl source} ever built, can be expected to make substantial improvements
over our present knowledge of those parameters.

The following sections are therefore dedicated to
illustrating the details (and the limitations) of present--day
information on low--energy kaon--nucleon
physics, spotlighting those points which still await being clarified,
and where DA$\Phi$NE can be expected to improve. Being the
phenomenology in this case more complex than in the (strictly related)
pion--nucleon one, we shall start almost from scratch.

We shall also take the liberty of not going into the details of models,
in particular for the spectroscopic classification of the
$J^P = {1\over2}^-$, $S$--wave resonance $\Lambda(1405)$: data are still
so scarce, after more than two decades of studies,
that any interpretation of such a state is to be
regarded as purely conjectural$^{4,5}$.

\vglue 1.0truecm
\leftline{\bf 2. Amplitude formalism for two--body $KN$ and $\overline{K}N$
interactions. }
\vglue 0.3truecm

Any $a_1(0^-,q) + B_1({1\over2}^+,p) \to a_2(0^-,q') + B_2({1\over2}^+,p')$
process is most economically described in the centre--of--mass (c.m.) frame
by two amplitudes, $G(w,\theta)$ and $H(w,\theta)$, when the T--matrix element
$T_{\alpha\beta}$ is expressed in terms of the two--component Pauli spinors
$\chi_\alpha$ and $\chi_\beta$ (respectively for the final and initial
${1\over2}^+$ baryons) as $T_{\alpha\beta} = \chi_\alpha^{\dag} \hbox{\bf T}
\chi_\beta$, where
$$ \hbox{\bf T} = G(w,\theta) \times \hbox{\bf I} + i H(w,\theta) \times
(\vec\sigma \cdot \hat n) \eqno(1) $$
and $\hat n$ defines the normal to the scattering plane$^5$.

These c.m. amplitudes have a simple expansion in the partial waves
$T_{\ell\pm}(w) = (\eta_{\ell\pm} e^{2 i \delta_{\ell\pm}} - 1) / 2 i q$,
given by
$$ G_N(w,\theta) = \sum_{\ell=0}^\infty [ (\ell + 1) T_{\ell+}(w) + \ell \
T_{\ell-}(w) ] P_\ell(\cos\theta) \eqno(2) $$
$$ H_N(w,\theta) = \sum_{\ell=1}^\infty [ T_{\ell+}(w) - T_{\ell-}(w) ]
P'_\ell(\cos\theta)\ , \eqno(3) $$
where the subscript $N$ indicates that only the {\sl purely nuclear} part of
the interaction has been considered.

To describe adequately the data, the amplitudes must also include
electromagnetism and can be rewritten as
$$ G(w,\theta) = \tilde G_N(w,\theta) + G_C(w,\theta) \eqno(4)$$
$$ H(w,\theta) = \tilde H_N(w,\theta) + H_C(w,\theta) \ ,\eqno(5)$$
where the tilded nuclear amplitudes differ from the untilded ones only in the
(generally complex) Coulomb shifts $\sigma_{\ell\pm}^{\rm in(fin)}$ having
been applied to each partial wave $T_{\ell\pm}$, namely when
$$ T_{\ell\pm}\ \to\ \tilde T_{\ell\pm} = e^{i \sigma_{\ell\pm}^{\rm
in}} T_{\ell\pm}(w) e^{i \sigma_{\ell\pm}^{\rm fin}} \ . \eqno(6)$$

The one--photon--exchange amplitudes $G_C$ and $H_C$ (of course
absent for charge-- and/or strangeness--exchange processes, but present at
$t \neq 0$ for $K_S$ regeneration, which can also go via one--photon
exchange) can be expressed in terms of the Dirac nucleon form factors
as$^{6,7}$ ($M$ and $m$ indicate respectively the baryon and meson masses
in the inital state: for final states all quantities will be primed)
$$ G_C(w,\theta) = \pm e^{\pm i \phi_C} \cdot \{ ( {{2 q \gamma} \over t}
+ {\alpha
\over {2 w}} {{w + M} \over {E + M}} ) \cdot F_1(t) + [ w - M + {t \over {4 (E
+ M)}} ] \cdot {{\alpha F_2(t)} \over {2 w M}} \} \cdot F_K(t) \eqno(7) $$
and
$$ H_C(w,\theta) = \pm {{\alpha F_K(t)} \over {2 w \tan{1\over2}\theta}} \cdot
\{ {{w + M} \over {E + M}} \cdot F_1(t) + [w + {t \over {4 (E + M)}}] \cdot
{{F_2(t)} \over M} \} \eqno(8) $$
for the interactions of (respectively) $K^\pm$ with nucleons, while for
$K_S^0$ regeneration one has to change the sign of the isovector part
$F_K^V(t)$ of the kaon form factor $F_K(t) = {1\over 2} [F_K^S(t) \pm F_K^V(t)
]$, the plus sign holding for charged kaons, the minus for the neutral ones.
Here $\gamma = \alpha \cdot (w^2 - M^2 - m^2) / 2 q w$ and the Coulomb phase
$\phi_C$ is defined as
$$ \phi_C = - \gamma \log(\sin^2{1\over2}\theta) + \gamma \cdot
\int_{-4q^2}^0 {{dt} \over t} \cdot [1 - F_K(t) F_1(t)] \eqno(9) $$
for charged kaons scattering on protons, while it reduces to
$$ \phi_C = -\gamma \int_{-4q^2}^0 {{dt}\over t} F_K(t) F_1(t) \eqno(9') $$
for processes involving $K^0$'s and/or neutrons.

We have denoted with $w$ and $\theta$ respectively the total energy
and the scattering angle in the c.m. frame, $q = [{1\over4} w^2 - {1\over2}
(M^2 + m^2) + (M^2 - m^2)^2 / 4 w^2]^{1/2}$ is the c.m. momentum (in the
initial state: for inelastic processes, including charge exchange, we shall
indicate final--state kinematical quantities with primes), $E$ the total
energy of the baryon in the c.m. frame, $E = (w^2 + M^2 - m^2)/2w$, and $t$
the squared momentum--transfer, $t = M^2 + M'^2 - 2 E E' + 2 q q' \cos\theta$.
We shall also use the laboratory--frame, initial--meson momentum $k =
{1\over2} (\omega^2 - m^2)^{1/2}$ and energy $\omega$, related to the
c.m. total energy via $\omega = (w^2 - M^2 - m^2)/2M$, and, besides $t$,
the two other
Mandelstam variables $s = w^2$ and $u$, the square of the c.m. total energy
for the crossed channel $\bar a_2(0^-) + B_1({1\over2}^+) \to \bar a_1(0^-)
+ B_2({1\over2}^+)$, obeying {\sl on the mass shell} the indentity $s
+ t + u = M^2 + M'^2 + m^2 + m'^2$.

In terms of the amplitudes $G$ and $H$ the c.m. differential cross
sections for an unpolarized target (surely the case for experiments to be
carried on at DA$\Phi$NE) take the simple form
$$ {{d\sigma} \over {d\Omega}} = {1\over2} \sum_{\alpha,\beta} \vert
T_{\alpha\beta} \vert^2 = \vert G \vert^2 + \vert H \vert^2 \ . \eqno(10) $$

The other observables possibly accessible at DA$\Phi$NE, in the
strangeness--exchange processes $\overline{K}N\to\pi\Lambda$ and
$\overline{K}N\to\pi\Sigma$, are the polarizations $P_Y$ ($Y=\Lambda$ or
$\Sigma$) of the final hyperons, measurable through the asymmetries $\alpha$
of their weak nonleptonic decays $\Lambda\to\pi^-p$ or $\pi^0n$, for both of
which we have an asymmetry $\alpha\simeq+0.64$, and $\Sigma^+\to\pi^0p$ for
which the asymmetry is $\alpha\simeq-0.98$, while there is very little
chance to be able to use the neutron decay modes $\Sigma^\pm\to\pi^\pm n$,
which have the very small asymmetries $\alpha\simeq\pm0.068$; we have for
these quantities
$$ P_Y\cdot({{d\sigma}\over{d\Omega}}) = 2\ \hbox{\rm Im}\ (GH^*)\ .
\eqno(11) $$

Note that, for an $(S+P)$--wave parametrization (fully adequate at such low
momenta), while the {\sl integrated} cross sections depend only {\sl
quadratically} on the $P$--waves, both the first Legendre coefficients of the
differential cross sections
$$ L_1 = {1\over2}\int_{-1}^{+1}\cos\theta\ ({{d\sigma}\over{d\Omega}})\
d\cos\theta = {2\over3}\ \hbox{\rm Re}\ [T_{0+}(2T_{1+}+T_{1-})^*+\dots]
\eqno(12)$$
and the polarizations
$$ P_Y\cdot({{d\sigma}\over{d\Omega}}) = 2\ \hbox{\rm Im}\ [T_{0+}(T_{1+}-
T_{1-})^*+3T_{1-}T_{1+}^*\cos\theta+\dots]\sin\theta
\eqno(13) $$
are essentially {\sl linear} in the small $P$--wave contributions, and give
two complementary pieces of information on these latter.
It is perhaps not useless to remind
the reader that the low statistics of the experiments, performed {\sl only} up
to the late seventies, have not been enough to determine any of these
parameters, yielding only very inaccurate (and utterly useless)
determinations$^2$ of $L_1$ for the charged--hyperon production channels.

We shall now devote the last part of this section to show explicitly why
this absence of {\sl direct} information on the low--energy behaviour of the
$P$--waves has been a serious shortcoming for $\overline{K} N$ amplitude
analyses. Remember that we know, from {\sl production} experiments, that the
$I = 1$, $S = -1$ $T_{1+}$ partial wave resonates {\sl below} threshold at a
c.m. energy around $w = 1384\ MeV$, the mass of the neutral, isovector
member of the $J^P={3\over2}^+$ decuplet$^5$, but beyond this piece
of information $P$--waves
are practically unmeasured up to momenta above $\simeq 500\ MeV/c$.

One has to turn from the Pauli amplitudes $G$ and $H$ to the invariant
amplitudes $A(s,t)$ and $B(s,t)$, defined in term of four--component Dirac
spinors as
$$ 2\pi w \ T_{\alpha\beta} = \bar u_\alpha(p') [A(s,t) + B(s,t)\cdot
\gamma^\mu Q_\mu ] u_\beta(p) \ , \eqno(14) $$
where $Q = {1\over2} (q + q')$, the average between incoming-- and
outgoing--meson c.m. four--momenta: these amplitudes obey simple
crossing relations and are free of kinematical
singularities, so that they are the ones to be used, rather than $G$ and $H$,
for any analytical extrapolation purpose; it is also customary to use the
combination $D(\nu,t) = A(\nu,t) + \nu \cdot B(\nu,t)$, where $\nu = (s - u)
/ 2(MM')^{1/2}$, which has the same properties as $A(\nu,t)$ under crossing,
and furthermore, for elastic scattering, obeys the optical theorem
in the simple form
$$ \hbox{\rm Im} \ D(\nu,t=0) = k \cdot \sigma_{tot} \ , \eqno(15) $$
where of course all electromagnetic effects must be subtracted on {\sl both}
sides.

One can rewrite $A$ and $B$ in terms of $G$ and $H$, and thus reexpress them
through the partial waves $T_{\ell\pm}$, by projecting eq. (14) on the
different spin states (polarized {\sl perpendicularly} to the scattering
plane) and obtain in the most general kinematics
\vfill\eject
$$ A(\nu,t) = {{4 \pi} \over {(E + M)^{1/2} (E' + M')^{1/2}}} \{ [w +
{1\over2} (M + M')] G(w,\theta) + $$
$$+ [(E + M) (E' + M') \{ w - {1\over2} (M + M') \} + \{ {1\over2} t + E E' -
{1\over2} (M^2 + M'^2) \} \{ w + {1\over2} (M + M') \} ] \cdot$$
$$\cdot {{H(w,\theta)} \over {q q' \sin\theta}} \} \ , \eqno(16)$$
\vglue 0.3truecm
and
$$ B(\nu,t) = {{4 \pi} \over {(E + M)^{1/2} (E' + M')^{1/2}}} \{ G(w,\theta)
- $$
$$- [(E + M) (E' + M') - {1\over2} t - E E' + {1\over2} (M^2 + M'^2)]
{{H(w,\theta)} \over {q q' \sin\theta}} \} \ . \eqno(17)$$
\vglue 0.3truecm

Considering for sake of simplicity forward elastic scattering only, the
amplitudes become, leaving out $D$-- and higher waves,
$$ D(\nu,0) = {{4 \pi w} \over M} [T_{0+} + 2 T_{1+} + T_{1-} + \dots ]
\eqno(18) $$
and
$$ B(\nu,0) = {{4 \pi w} \over {M q^2}} [(E - M) T_{0+} - 2 (2M - E) T_{1+} +
(E + M) T_{1-} + \dots ] \ ; \eqno(19) $$
introducing the (complex) scattering lengths $a_{\ell\pm}$ and (complex)
effective ranges $r_{\ell\pm}$ one can expand up to $O(q^2)$ the partial
waves close to threshold, and obtain for the forward $D$ amplitudes
$$ D(q,0) = 4 \pi (1 + {m\over M}) \{ a_{0+} + i a_{0+}^2 q + [2 a_{1+} +
a_{1-} - (a_{0+} + {1\over2} r_{0+}) a_{0+}^2 - {a_{0+} \over {2 M m}}] q^2
+ \dots \} \ , \eqno(20) $$
dominated by the $S$--waves, while for the $B$ amplitudes the same
approximations give
$$ B(q,0) = {{2 \pi} \over M} (1 + {m \over M}) [a_{0+} - 4 M^2 (a_{1+} -
a_{1-}) + i a_{0+}^2 q + \dots] \ , \eqno(21) $$
where the factor $4 M^2 \simeq 90\ fm^{-2}$ enhances considerably the
contributions by the low--energy P--waves (virtually unkown), rendering
practically useless the unsubtracted dispersion relation for the better
converging $B$ amplitudes, so important for the $\pi N$ case in fixing
accurately the values of the coupling constant $f^2$ and of the $S$--wave
scattering lengths$^6$.

\vglue 1.0truecm
\leftline{\bf 3. The Coulomb corrections and the kaonic hydrogen ``puzzle''. }
\vglue 0.3truecm

The Coulomb shifts $\sigma_{\ell\pm}$ can be separated into the real, purely
Coulomb phases $\sigma_{\ell}$ and the {\sl complex} Coulomb corrections
$\Delta_{\ell\pm}$,
$$\sigma_{\ell\pm} = \sigma_\ell + \Delta_{\ell\pm} \ , \eqno(22)$$
where $\sigma_\ell = \hbox{\rm arg} \Gamma (\ell + 1 + i \gamma)$ to lowest
order in $\alpha$, and the Coulomb corrections $\Delta_{\ell\pm}$ have to be
explicitly computed from a first--order ``ansatz'' on the purely nuclear
interaction. They have been computed in a dispersion--relation formalism by
Tromborg, Waldenstr{\o}m and {\O}verb{\o}$^8$ for the $\pi N$ case: the same
formalism$^7$ could in principle be extended (but has not been up to now) also
to the $KN$ and $\overline{K}N$ ones.

Since this method (as the other methods as well) uses a ``reference'' strong
interaction, minor corrections remain to be applied to the {\sl observed}
phases $\delta_{\ell\pm}$ and elasticities $\eta_{\ell\pm}$, to extract their
{\sl purely nuclear} parts$^6$; a way of removing them efficiently has been
devised by the Karlsruhe--Helsinki group: it consists in starting from a
(preliminary) set of phase shifts, (i) calculating from them the corrections
in the above--mentioned dispersive formalism$^{7,8}$, then (ii) the changes in
the observables brought about by these latter, and finally (iii) correcting
the data for these effects and (iv) starting the phase--shift analysis all
over again, this time from the ``corrected data''. This procedure has turned
out to be both self--consistent and fast$^6$.

Of course, the accuracy of a dispersive approach is limited by the {\sl
overall} accuracy of the experimental data, from which the ``reference''
interaction is extracted, and particularly by that of the low--energy ones,
which in the $\overline{K}N$ interaction is dominated by the very low
statistics of the most accurate -- and recent -- experiment$^2$.

As an alternative to the above dispersive approach$^{7,8}$, one can describe
the scattering and production processes by either
a Schr\"odinger$^{9,10}$ or a
Klein--Gordon$^{11,12}$ multichannel wave equation, with both Coulomb and
strong--interaction potentials; such formalisms have the advantage, when
continuing the c.m. momentum $q$ of the $K^-p$ system from the real to the
imaginary axis, of predicting at the same time the hadronic level shifts and
widths for the kaonic--hydrogen atomic levels, present as an infinite number
of poles just below the elastic threshold in every partial wave, having the
threshold as an accumulation point.

Such an approach is preferable when working at very low momenta, in absence of
previous experimental information of comparable quality, which is the case for
the foreseeable $KN$ and $\overline{K}N$ experiments to be carried out at
DA$\Phi$NE. We shall in the following discuss in detail only the relatively
less know Klein--Gordon formalism, leaving out the Schr\"odinger one, whose
details are easily workable out, following e.g. the papers listed in reference
(10) as a guideline.

The advantage of the Klein--Gordon formaliam over the Schr\"odinger one lies
mainly in the fact that the fine structure of mesic atoms is clearly
(and directly)
produced by the former but not by the latter: one would thus be tempted to use
it also to calculate the Coulomb modifications to the scattering amplitudes,
particularly in the low--energy region considered here, where both formalisms
give reasonable approximations to the true kinematics of the two--particle
system. We have indeed, for the c.m. kinetic energy $T_{c.m.} = [M^2 +
q^2]^{1\over2} + [m^2 + q^2]^{1\over2} - (M + m)$, the approximations $T_S =
q^2/2\mu$  -- with the reduced mass $\mu = Mm/(M+m)$ -- for the Schr\"odinger
equation, and $T_{KG} = E_{KG} - \mu = [\mu^2 + q^2]^{1\over2} - \mu$ for the
Klein--Gordon one, both differing from each other and from $T_{c.m.}$ at
O($q^4$) only.

However, to include an interaction with all the good symmetry properties
expected for pseudoscalar mesons interacting with a baryonic ``source'', a
Klein--Gordon equation has to possess at least a four--vector and a scalar
term (being concerned mostly with the $S$--waves, we shall neglect for the
moment the tensor part of the interaction), of the form (in the ``static
limit'')
$$\{\nabla^2 - \mu^2 - 2\mu\ S(r) - [E_{KG} + U(r) - e\varphi_C(r)]^2\}\
\Psi({\hbox{\bf r}})\ =\ 0\ , \eqno(23)$$
where the {\sl effective} potentials $S(r)$ and $U(r)$ (real in the elastic,
single--channel case) are respectively even and odd under $C$--parity
conjugation of the meson fields (in a multi--channel case, such as the one we
are studying, $S(r)$ and $U(r)$ will be real, nondiagonal matrices, and
$E_{KG}$ and $\mu$ diagonal ones, of dimension $6 \times 6$ for $K^-p$ and
$5 \times 5$ for $\overline{K}^0p$, and the product $2\mu S(r)$ will have to
be replaced by the anticommutator $\{\mu,S(r)\}$): they can be separated only
if data on {\sl both} the $s$-- and $u$--channel reactions are available {\sl
at the same c.m. energy}, which is clearly {\sl not} the case for the
pion--hyperon channels, since the processes $\pi p \to KY$ can be accessed
only at c.m. energies above $m_K + M_Y$, well above the $\overline{K}N$ c.m.
energy available at DA$\Phi$NE, below 1442 $MeV$ for $K^-p$ (and of only 1444
$MeV$ for $K_L^0p$).

One has therefore to abandon the real potentials for complex ones, reducing
oneself to treat with the Klein--Gordon equation (23) only the two, coupled
$K^-p$ and $\overline{K}^0n$ channels: however, one ambiguity is still left,
in the separation of the absorption effects from the neglected $\pi\Lambda$
and $\pi\Sigma$ channels between the two potentials $S(r)$ and $U(r)$. This
ambiguity could be resolved$^{12}$, under the {\sl reasonable}
assumption of an energy
dependence of the potentials$^{13}$ as gentle as expected from forward
dispersion relations and the absence of any dynamical effect apart from the
$\Lambda(1405)$ $S$--wave resonance below threshold$^{4,5}$,
if the energy shift
and width of the ground state for the $K^-p$ atomic system had been actually
measured: the three experiments performed in the late seventies and early
eighties on this system$^{14,15}$ have been widely quoted as to report {\sl
attractive} hadronic shifts (and small widths), contrary to all {\sl
reasonable} expectations$^{16,17}$, but in fact none of the three has produced
convincing evidence for the identification of the X--ray transitions they
claim having observed as lines belonging to the K--series of the $K^-p$ atomic
system. The only experiment having an acceptable signal--to--background ratio
(the first one conducted by Davies {\sl et al.}$^{14}$ in the same beam--line
of the TST Collaboration experiment$^2$ at NIMROD), observed only {\sl one}
clear transition, which could well have been the $K^-p$ K$_\beta$ line and
{\sl not} the K$_\alpha$ they claimed it to be (see the comments about this
problem in both reviews listed in reference (16)).
For the interpretation of the $K^-p$ X--ray
lines the evaluation of their intensities is essential: a recent cascade
calculation has been performed by Reifenr\"other and Klempt$^{18}$. A new
experiment, which expects to achieve a much better signal--to--background
ratio, is presently under way at KEK$^{19}$, and studies on the possibility of
a kaonic hydrogen experiment at DA$\Phi$NE must be seriously considered, for
this measurement will be complementary to, if not substitutive of,
a conventional beam--target experiment
such as the KEK one, due to the completely different nature in the possible
X-ray backgrounds.

The above ambiguity is not as serious an obstacle as it might appear at first
glance: the ``strong potentials'' $S(r)$ and $U(r)$ are usually assumed to
have a simple, predetermined $r$--dependence (with Rasche
and Woolcock$^{10}$
we prefer square wells$^{12}$, $S(r) = S_0\ \Theta(r_0 - r)$ and $U(r) = U_0\
\Theta(r_0 - r)$, since these yield very simple, analytic ``inner'' and
``outer'' solutions, i.e. respectively Bessel and Kummer (or Whittaker)
functions, once $\varphi_C(r)$ is substituted by its average value inside the
square well, and vacuum polarization is neglected at this stage, and later
treated as a small perturbation, though this requires some particular
care$^{20}$ in normalising the Klein--Gordon wavefunctions, with respect to
the Schr\"odinger ones), and the ambiguity reduces then to a {\sl single} free
parameter, e.g. the ratio $1 \ge \rho = {\hbox{\rm Im}} S_0/ ({\hbox{\rm Im}}
S_0 + {\hbox{\rm Im}} U_0) \ge 0$ (since unitarity dictates both kind of
potentials, when absorptive, to have a {\sl negative} imaginary part). The
calculation can now go on along the same lines as the Helsinki--Karlsruhe one
({\sl modulo} the different algorithms): one can start from an ``initial'' set
of phases, (i) turn them into a pair of potentials $S$, $U$, (ii) compute the
new corrections, (iii) use these latter to extract new phases, and (iv) start
again from the point (i) till input and output differ by less than a small,
pre--fixed amount.

The procedure relies, rather than on cumbersome integrations (as in the
dispersive approach), on well known analytic functions, most of which if not
all already available in software packages, easy to adapt to the case under
study; the same software, after extrapolation of the potentials to threshold
through a reasonably small correction ($q$ in the region covered by
DA$\Phi$NE's $K^-$'s goes from about 59 to about 78 $MeV/c$ only), since it
contains automatically the Deser--Goldberger--Baumann--Thirring formula for
the hadronic shifts and widths$^{12,21}$, can also produce, {\sl mutatis
mutandis}, accurate predictions for shifts and widths of the $K^- p$ 1$s$ (and
$Np$, once the $P$--waves are determined as well) atomic states, and therefore
for the {\sl whole} K--series, for any chosen set of the only two free
parameters left, i.e. the ratio $\rho$ and the square--well radius $r_0$, to
be varied at most from once to twice the r.m.s. charge radius of the proton
(or up and down a $30\%$ around $1 \ fm$).

\vglue 1.0truecm
\leftline{\bf 4. Open channels and baryon spectroscopy at DA$\Phi$NE. }
\vglue 0.3truecm

As mentioned above, in the momentum region which could be explored by the
kaons coming from the decays of a $\phi$--resonance formed at rest in an $e^+
e^-$ collision, we have only data from low--statistics experiments, mostly
hydrogen bubble--chamber ones on $K^- p$ (and $K^-d$) interactions$^{1,2}$
(dating from the early sixties trough the late seventies), plus scant data
from $K^0_L$ interactions and $K^0_S$ regeneration, mostly on hydrogen$^3$.

The channels, open at a laboratory energy $\omega = {1\over2} m_\phi$ (for
$K^\pm$'s to obtain the exact value of $\omega$ one has to include their
energy losses through ionization as well), are tabulated below for
interactions with {\sl free} protons and neutrons, together with their
threshold energies $w_{thr}$ (in $MeV$), strangeness and isospin(s).
We do not list $K^+$--initiated processes, which are (apart from charge
exchange) purely elastic in this energy region.

For interactions in hydrogen, the c.m. energy available for each final
state is limited by momentum conservation to the initial total c.m. energy,
equal (neglecting energy losses) to $w = (M_p^2 + m_K^2 + M_p m_\phi)^{1/2}$,
or $1442.4\ MeV$ for incident $K^\pm$'s and $1443.8\ MeV$ for incident
$K^0_L$'s. Energy losses for charged kaons can be exploited (using the inner
parts of the detector as a ``moderator'') to explore $K^-p$ interactions in a
{\sl limited} momentum range, possibly down to and below
the charge--exchange threshold
at $w = 1437.2\ MeV$, corresponding to a $K^-$ laboratory momentum of about
$90\ MeV/c$.

For interactions in deuterium (or in heavier nuclei), momentum can be carried
away by ``spectator'' nucleons, and one can explore each inelastic channel
from the highest available energy down to threshold. The possibility of
reaching energies below the $\overline{K}N$ threshold is particularly
desirable, since the $\overline{K}N$ unphysical region contains two
resonances$^{4,5}$, the $I = 0$,
$S$--wave $\Lambda(1405)$ and the $I = 1$, $J^P = {3\over2}^+$
$P$--wave $\Sigma(1385)$, observed mostly in production
experiments (and, in the first case, with very limited statistics$^{22}$), so
that the information on their couplings to the $\overline{K}N$ channel relies
{\sl entirely} on extrapolations below threshold of the analyses of the
low--energy data. The coupling of the $\Sigma(1385)$ to the $\overline{K}N$
channel, for instance, can at present be determined only via forward
dispersion relations involving the {\sl total sum} of data collected at $t
\simeq 0$, but with uncertainties which are, {\sl at their best}, still of the
order of 50\% of the value expected from flavour--$SU(3)$ symmetry$^{23}$; as
for the $\Lambda(1405)$, even its spectroscopic classification is still an
open problem, {\sl vis--\`a--vis} the paucity and (lack of) quality of the
{\sl best available} data$^{4,24}$.

\vglue 1.0truecm
\centerline{\bf Table I}
\vglue 0.3truecm
\hrule
\vglue 0.3truecm
$$\vbox{\halign{#\hfil&\qquad\qquad\hfil#\hfil&\qquad\qquad\hfil#\hfil&\qquad
\qquad\hfil#\hfil\cr
\qquad \sl Channel & $w_{thr}/MeV$ & $S$ & $I$ \cr
$K^-p,K^0_Ln\ \to\ \pi^0\Lambda$ & 1250.6 & --1 & 1 \cr
$K^-p,K^0_Ln\ \to\ \pi^0\Sigma^0$ & 1327.5 & --1 & 0 \cr
$K^-p,K^0_Ln\ \to\ \pi^-\Sigma^+$ & 1328.9 & --1 & 0,1 \cr
$K^-p,K^0_Ln\ \to\ \pi^+\Sigma^-$ & 1337.0 & --1 & 0,1 \cr
$K^-p,K^0_Ln\ \to\ \pi^0\pi^0\Lambda$ & 1385.6 & --1 & 0 \cr
$K^-p,K^0_Ln\ \to\ \pi^+\pi^-\Lambda$ & 1394.8 & --1 & 0,1 \cr
$K^-p,K^0_Ln\ \to\ K^-p$ & 1431.9 & --1 & 0,1 \cr
$K^-p,K^0_Ln\ \to\ K^0_Sn$ & 1437.2 & --1 & 0,1 \cr
\qquad''\qquad'' & '' & +1 & 1$^{\dag}$ \cr}} $$
\hrule
\vglue 0.3truecm
\centerline{\dag) This amplitude only appears in the regeneration process
$K^0_Ln \to K^0_Sn$.}

\vglue 1.5truecm
\centerline{\bf Table II}
\vglue 0.3truecm
\hrule
\vglue 0.3truecm
$$\vbox{\halign{#\hfil&\qquad\qquad\hfil#\hfil&\qquad\qquad\hfil#\hfil&\qquad
\qquad\hfil#\hfil\cr
\quad\sl Channel & $w_{thr}/MeV$ & $S$ & $I$ \cr
$K^-n\ \to\ \pi^-\Lambda$ & 1255.2 & --1 & 1 \cr
$K^-n\ \to\ \pi^-\Sigma^0$ & 1332.1 & --1 & 1 \cr
$K^-n\ \to\ \pi^0\Sigma^-$ & 1332.1 & --1 & 1 \cr
$K^-n\ \to\ \pi^0\pi^-\Lambda$ & 1388.2 & --1 & 1 \cr
$K^-n\ \to\ K^-n$ & 1433.2 & --1 & 1 \cr}}$$
\hrule\vfill\eject

\centerline{\bf Table III}
\vglue 0.3truecm
\hrule
\vglue 0.3truecm
$$\vbox{\halign{#\hfil&\qquad\qquad\hfil#\hfil&\qquad\qquad\hfil#\hfil&\qquad
\qquad\hfil#\hfil\cr
\quad \sl Channel& $w_{thr}/MeV$ & $S$ & $I$ \cr
$K^0_Lp\ \to\ \pi^+\Lambda$ & 1255.2 & --1 & 1 \cr
$K^0_Lp\ \to\ \pi^0\Sigma^+$ & 1324.3 & --1 & 1 \cr
$K^0_Lp\ \to\ \pi^+\Sigma^0$ & 1332.1 & --1 & 1 \cr
$K^0_Lp\ \to\ \pi^0\pi^+\Lambda$ & 1388.2 & --1 & 1 \cr
$K^0_Lp\ \to\ K^+n$ & 1433.2 & +1 & 0,1 \cr
$K^0_Lp\ \to\ K^0_Sp$ & 1435.9 & +1 & 0,1 \cr
\quad''\qquad'' & '' & --1 & 1 \cr}}$$
\hrule
\vglue 1.0truecm

A {\sl formation} experiment on {\sl bound} nucleons in an (almost) $4\pi$
apparatus with good efficiency and resolution for low--momentum $\gamma$'s
(such as KLOE) can reconstruct and measure a channel such as $K^-p \to
\pi^0\Sigma^0$ (only {\sl above} the $\overline{K}N$ threshold) or $K^-d \to
\pi^0\Sigma^0n_s$ (both {\sl above} and {\sl below} threshold), which is pure
$I = 0$: up to now all analyses on the $\Lambda(1405)$ have been limited to
charged channels$^{22}$, being thus forced to assume the $I = 1$ contamination
in their samples to be either negligible or smooth and not interfering with
the resonance signal (remember that it is common knowledge$^5$ that there is a
$P$--wave resonance in the $I = 1$, $\pi^\pm\Sigma^\mp$ channels at 1384
$MeV$!). This situation is particularly unsatisfactory, in view of the fact
that the various spectroscopic models proposed for the classification of the
$\Lambda(1405)$ differ mostly in the detailed resonance {\sl shape}, rather
than in its couplings$^{24}$: now, it is precisely the shape which could be
drastically changed even by a moderate amount of interference with an $I = 1$
``background''. Note also that, having in the same apparatus, and at almost
the same energy {\sl tagged} $K^-$ and $K^0_L$ produced at the same point, one
can separate $I=0$ and $I=1$ channels with a minimum of systematic
uncertainties, by measuring all channels $K^0_Lp \to \pi^0\Sigma^+$,
$\pi^+\Sigma^0$ and $K^-p \to \pi^-\Sigma^+$, $\pi^+\Sigma^-$, besides, of
course, the above--mentioned, pure $I=0$, $K^-p \to \pi^0\Sigma^0$ one.

Another class of inelastic processes, which are expected to be produced
(even if at a much smaller rate) by DA$\Phi$NE's $\overline{K}$'s, is
radiative capture, leading in hydrogen to the final states $\gamma\Lambda$
and $\gamma\Sigma^0$ for incident $K^-$'s, and, for incident $K^0_L$'s, to the
final state $\gamma\Sigma^+$: in deuterium, one expects to observe the capture
processes by neutrons, $K^-d \to \gamma\Sigma^-p_s$ and $K_Ld \to
\gamma\Sigma^0p_s$, $\gamma\Lambda p_s$ as well. Observation of these
processes has been limited up to now to searches for photons emitted after
capture of $K^-$'s stopped in liquid hydrogen (and deuterium): but in these
experiments the spectra are dominated by photons from unreconstructed $\pi^0$
and $\Sigma^0$ decays$^{25}$. This poses serious difficulties already at the
level of separation of signals from background, since (in $K^-p$ capture at
rest) only the photon line from the $\gamma\Lambda$ final state falls {\sl
just above} the endpoint of the photons from decays of the $\pi^0$'s in the
$\pi^0\Lambda$ final state, while that from $\gamma\Sigma^0$ falls right on
top of this latter: indeed, these experiments were able to produce, within
quite large errors, only an estimate of the respective branching ratios.

The $4\pi$ geometry possible at DA$\Phi$NE, combined with the ``transparency''
of a KLOE--like apparatus, its high efficiency for photon detection and its
good resolution for spatial reconstruction of the events, should make possible
the full identification of the final states and therefore the measurement of
the absolute cross sections for these processes, although in flight and not
at rest.

This difference can be appreciated when comparing with theoretical
predictions: the main contributions to radiative captures are commonly
thought to come from radiative decays of resonant levels in the
$\overline{K}N$ system$^{26}$, while the total hyperon production cross
section is expected to come from both resonant and non--resonant intermediate
states. An estimate of the branching ratios would therefore be quite sensitive
to the latter, while a prediction of the absolute cross sections should not.

Data$^{25}$ are presently indicating branching ratios around $0.9 \times
10^{-3}$ for $K^-p \to \gamma\Lambda$ and $1.4\times 10^{-3}$ for $K^-p \to
\gamma\Sigma^0$, with errors of the order of 15\% on both rates: most
theoretical models$^{27}$ tend to give the first rate larger than the
second, with both values consistently higher than the observed ones. Only a
cloudy--bag--model estimate$^{28}$ exhibits the trend appearing (although
only at a $2\sigma$ level, and therefore waiting for confirmation by better
data) from the first experimental determinations, but this is the only respect
in which this model agrees with the data, still giving branching ratios larger
than observations by a factor two$^{29}$.

Data are also interpretable in terms of $\Lambda(1405)$ electromagnetic
transition moments$^{26}$: this interpretation of measurements taken at a
single energy, or over a limited interval, is clearly subject to the effect
of the interference between this state and all other contributions, such as
the $\Lambda$-- and $\Sigma$--hyperon poles and other resonant states such
as the $\Sigma(1385)$ and the $\Lambda(1520)$, not to mention $t$--channel
exchanges (since at least $K$--exchange has to be included, to ensure gauge
invariance of the Born approximation). An extraction of the $\Lambda(1405)$
moments, relatively freer of these uncertainties, requires measurements of
the final states $\gamma\Lambda$ and $\gamma\Sigma$ (if possible, in
different charge states) over the {\sl unphysical region}, using (gaseous)
deuterium or helium as a ``target''. Rates are expected to be only of the
order of $10^4\ events/y$, but it must be kept in mind that such a low rate
(by DA$\Phi$NE's standards) corresponds already to statistics {\sl two} orders
of magnitude above those of the {\sl best} experiment performed till now on
the shape of the $\Lambda(1405) \to \pi\Sigma$ decay spectrum$^{22}$.

\vglue 1.0truecm
\leftline{\bf 5. The K-matrix (or M-matrix) formalism. }
\vglue 0.3truecm

An adequate description of the low--energy $\overline{K}N$ partial waves must
couple at least the dominant, two--body inelastic channels to each other and
to the elastic one; the three--body channel $\pi\pi\Lambda$ is expected to be
suppressed, for $J^P = {1\over2}^-$, by the angular momentum barrier, but it
could contribute appreciably to the $I=0$, $J^P={1\over2}^+$ $P$--wave, due
to the strong final--state interaction of two pions in an $I=0$ $S$--wave.
Note that {\sl most} bubble chamber experiments were unable to fully
reconstruct the events at the lowest momenta, and therefore often assumed all
{\sl directly} produced $\Lambda$'s to come from the $\pi\Lambda$ channel {\sl
alone}, neglecting altogether the small $\pi\pi\Lambda$ contribution.

The appropriate formalism is to introduce a K--matrix description (sometimes
it is convenient to use, instead of the K--matrix, its inverse, also known as
the M--matrix), defined in the isospin eigenchannel notation as
$$ \hbox{\bf K}_{\ell\pm}^{-1} = \hbox{\bf M}_{\ell\pm} =
\hbox{\bf T}_{\ell\pm}^{-1} + i\ \hbox{\bf Q}^{2\ell +1}\ , \eqno(24)$$
for both $I=0, 1$ $S$--waves (and perhaps also for the four $P$--waves as
well, or at least for the $I=1$, $T_1+$ wave, which resonates below
threshold). The K--matrices, assuming $SU(2)$ symmetry, describe the
$S$--wave data
at a given energy in terms of {\sl nine} real parameters (six for $I=1$ and
three for $I=0$), while the experimentally accessible processes can be
described, assuming pure $S$--waves in the same symmetry limit, by only {\sl
six} independent parameters, which can be chosen to be the two (complex)
amplitudes $A_0,\ A_1$ for the $\overline{K}N \to \overline{K}N$ channel, the
phase difference $\phi$ between the $I=0$ and $I=1$ $\pi\Sigma$ production
amplitudes, and the ratio $\epsilon$ between the $\pi\Lambda$ production cross
section and that for total hyperon production in an $I=1$ state$^{30}$.

Thus a single--energy measurement does not allow a complete determination of
the K--matrix elements at that energy. Using high--statistics measurements
at different momenta, and assuming either {\sl constant} K--matrices or (if
more complexity were needed) {\sl effective--range} M--matrices could
{\sl in principle} fully determine the matrix elements: but for this to be
possible one has to be able to subtract out the (small) $P$--wave
contributions to the integrated cross sections
$$ \sigma = 4\pi L_0 = 2\pi \int_{-1}^{+1} ({{d\sigma}\over{d\Omega}})\
d\cos\theta = 4\pi [ \vert T_{0+} \vert^2 + 2 \vert T_{1+}\vert^2 + \vert
T_{1-}\vert^2 + \dots ] \ , \eqno(25)$$
which could be obtained either from $L_1$ alone, for the elastic and
charge--exchange channels, or from both $L_1$ and $P_Y$, which give {\sl
complementary} information, for the hyperon production channels. None of these
quantities has been measured with the desirable accuracy up to now:
the TST Collaboration tried to extract
$L_1$ from some of their low--statistics data, and found results consistent
with the tail of the $\Sigma(1385)$ resonance in the $I=1$, $T_{1+}$ wave, but
also consistent (at the $2\sigma$ level) with zero within their obviously very
large errors$^2$. At the same level of accuracy, one should also be able to
isolate and separate out the $\pi\pi\Lambda$ channel contribution as well.

Remember that an accurate analysis has also to include the {\sl complete} e.m.
correct\-ions: up to now all $\overline{K}N$ analyses have relied on the
formula derived by Dalitz and Tuan$^{31}$ for a pure $S$--wave
scattering with a weak hadronic, short--range interaction, which is hardly the
case for the $\overline{K}N$ system around threshold.

To fix the {\sl redundant} K--matrix parameters, different authors have tried
different methods: some have used the data on the shape of the $\pi\Sigma$
spectrum from production experiments$^{32}$, others have constrained the
amplitudes in the unphysical region by imposing consistency with dispersion
relations for the amplitudes $D$ for both $K^\pm p$ and $K^\pm n$ forward
elastic scattering$^{33,34}$, relying on the existence of accurate data on the
total cross sections at higher energies. More recently, some attempts have
been made to combine both constraints into a ``global'' analysis, but with no
better results than each of them taken separately$^{35}$.

Unfortunately, neither of these methods has been very powerful, because of the
low statistics of the $\pi\Sigma$ production data on one side, and on the
other because of the need to use for the dispersion relations the often not
very accurate information (and particularly so for the $K^\pm n$ amplitudes)
on the real--to--imaginary--part ratios.

We list below (without errors, often meaningless since the parameters are
strongly correlated, and therefore not even quoted by some of the authors) the
constant K--matrices found by Chao {\sl et al.} using the first method$^{32}$
(which {\sl did not} include the TST Collaboration data), and the more complex
parametrization found by A.D. Martin using the second$^{34}$ (and including
the {\sl preliminary} TST data). Note that to describe the data for $I=0$ both
{\sl above} and {\sl below} threshold A.D. Martin was forced to introduce a
``constant--effective--range'' M--matrix, where ${\bf M}^{(0)} = ({\bf
K}^{(0)})^{-1} = {\bf A} + {\bf R} q^2$, with three more ``effective range''
parameters, so that to make the two analyses comparable we list separately his
{\sl threshold} K--matrix values.

The purpose of this table is to show that there is considerable uncertainty
even on the value of the $K^{(I)}_{NN}$ elements of the K--matrices (the real
parts of the corresponding scattering lengths): the data have been
re--analyzed by Dalitz {\sl et al.}$^{35}$, using {\sl both} sets of
constraints with different weigths and different parametrizations, and
yielding a variety of fits, all of them of about the same overall quality and
none of them improving very much over the above ones.

Just to highlight the difficulties met in describing the data (probably
plagued by inconsistencies between different experiments, and by
large systematic uncertainties), we point out
that A.D. Martin himself$^{34}$ found that including in his analysis a
$\Sigma(1385)$ resonance at the right position, with the width given by
the production experiments (and listed in the Particle Data Group tables$^5$)
and the coupling to the $\overline{K}N$ channel dictated by flavour--$SU(3)$
symmetry, was worsening rather than improving the fits obtained
{\sl neglecting} it altogether: his analysis therefore considers
the $\Sigma$ Born--term contribution a ``superposition'' of the former and
of that of the $P$--wave resonance, a rather unsavoury situation considering
the different $J^P$ quantum numbers of the two states, which may raise
questions about the applicability of his analysis away from $t \simeq 0$.
Note that a similar superposition
has to be considered in the $K^\pm p$ dispersion relations for the $\Sigma$--
and $\Lambda$--pole contributions, which can not be separated from each other
due to their being very close in the $\nu$--variable plane:
here however the two states contribute to the same partial wave, and the
$\Sigma$--pole can be independently extracted from $K^\pm n$ scattering
(or $K^0_S$ regeneration on protons) data$^{36}$.

In the analysis of the low--energy data collected
in the past on these processes, one of the main difficulties
comes from the large spread in momentum of the
typical low--energy kaon beams, for $K^\pm$'s because of the degrading in a
``moderator'' of the higher--energy beams needed to transport the kaons
away from their production target, for $K^0_L$'s because of the large
apertures needed to
achieve satisfactory rates in the targets (typically bubble chambers): this
made unrealistic the proposals (advanced from the early seventies) of better
determining the low--energy
K--matrices by studying the behaviour of the cross sections for
$K^-p$--initiated processes at the
$\overline{K}^0n$ charge--exchange threshold$^{37}$.
The high momentum resolution
available at DA$\Phi$NE will instead make such a goal a realistically
achievable one.
\vglue 0.8truecm
\centerline{\bf Table IV}
\vglue 0.3truecm
\hrule
$$\vbox{\halign{\hfil # \hfil & \quad \hfil # \hfil & \quad \hfil # \hfil
& \quad \hfil # \hfil \cr
\sl Chao et al. & & \sl A. D. Martin & \cr
& & & \cr
$K_{NN}^{(0)} = -1.56 fm$ & $A_{NN} = -0.07 fm^{-1}$ & $R_{NN} = +0.18 fm$
& $K_{NN}^{(0)}(0) = -1.65 fm$ \cr
$K_{N\Sigma}^{(0)} = -0.92 fm$ & $A_{N\Sigma} = -1.02 fm^{-1}$ & $R_{N\Sigma}
= +0.19 fm$ & $K_{N\Sigma}^{(0)}(0) = -0.87 fm$ \cr
$K_{\Sigma\Sigma}^{(0)} = +0.07 fm$ & $A_{\Sigma\Sigma} = +1.94 fm^{-1}$ &
$R_{\Sigma\Sigma} = -1.09 fm$ & $K_{\Sigma\Sigma}^{(0)}(0) = +0.06 fm$ \cr
& & & \cr
$K_{NN}^{(1)} = +0.76 fm$ & & &$K_{NN}^{(1)} = +1.07 fm$ \cr
$K_{N\Sigma}^{(1)} = -0.97 fm$ & & & $K_{N\Sigma}^{(1)} = -1.32 fm$ \cr
$K_{N\Lambda}^{(1)} = -0.66 fm$ & & & $K_{N\Lambda}^{(1)} = -0.30 fm$ \cr
$K_{\Sigma\Sigma}^{(1)} = +0.86 fm$ & & & $K_{\Sigma\Sigma}^{(1)} = +0.27 fm$
\cr
$K_{\Sigma\Lambda}^{(1)} = +0.51 fm$ & & & $K_{\Sigma\Lambda}^{(1)} = +1.54
fm$ \cr
$K_{\Lambda\Lambda}^{(1)} = +0.04 fm$ & & & $K_{\Lambda\Lambda}^{(1)} = -1.02
fm$ \cr}}$$
\hrule
\vglue 1.3truecm

In this case one can no longer assume
$SU(2)$ to be a good symmetry of the amplitudes: under the (reasonable)
assumption that the forces are still $SU(2)$--symmetric, one can however
still retain the previous K--matrix formalism, but one can no longer
decouple the different isospin eigenchannels$^{10}$.
Introducing the orthogonal matrix {\bf R}, which
transforms the six isospin eigenchannels for $\overline{K}N$ ($I=0, 1$),
$\pi\Lambda$ ($I=1$ only) and $\pi\Sigma$ ($I=0, 1, 2$) into the
six physical charge channels $K^-p$, $\overline{K}^0n$, $\pi^0\Lambda$,
$\pi^-\Sigma^+$, $\pi^0\Sigma^0$ and $\pi^+\Sigma^-$, and the diagonal matrix
{\bf Q}$_c$ of the c.m. momenta for these latter, one can rewrite
the T--matrix for the $S$--waves in the isospin--eigenchannel space as
$$ \hbox{\bf T}_I^{-1} = \hbox{\bf K}_I^{-1} - i \hbox{\bf R}^{-1} \hbox{\bf
Q}_c \hbox{\bf R} \ , \eqno(26)$$
where {\bf K}$_I$ is a box matrix with zero elements between channels of
different isospin, and {\bf R}$^{-1}$ {\bf Q}$_c$ {\bf R} is of course no
longer diagonal.

Apparently this involves one more parameter, since it
also contains the element $K_{\Sigma\Sigma}^{(2)}$: in practice, if one is
interested in the behaviour of the cross sections in the neighbourhood of
the $\overline{K}N$ charge--exchange threshold, one can take the
c.m. momenta in the three $\pi\Sigma$ channels as equal, so that the
$I=2$, $\pi\Sigma$ channel
decouples from the $I=0,\ 1$ ones, since the ``rotated'' matrix
{\bf R}$^{-1}${\bf Q}$_c${\bf R}
has now only two non--zero, off-diagonal elements, equal to
${1\over2} (q_0 - q_-)$ (where the subscripts refer to the kaon charges),
between the $I=0$ and $I=1$ $\overline{K}N$ channels,
the diagonal ones being almost the
same as in the {\sl fully} $SU(2)$--symmetric case, with only the substitution
to the $\overline{K}N$ channel momentum $q$ of the average over the two charge
states, ${1\over2} (q_0 + q_-)$. $K^{(2)}_{\Sigma\Sigma}$ would however be
important for describing accurate experiments on $\pi\Sigma$ and $\pi\Lambda$
mass spectra in the unphysical region below the $\overline{K}N$ threshold
without recourse to an $SU(2)$--symmetry limit: but the state--of--the--art
of our understanding of wave--functions, even for the lightest nuclei, is not
such as to make these isotopic--symmetry--breaking corrections relevant.

\vglue 1.0truecm
\leftline{\bf 5. Low--energy $K^+$ scattering is important, too. }
\vglue 0.3truecm

Better information on the $S=+1$ system is also essential in several cases. We
limit ourselves to mention only two of the problems coming to our mind.
Isospin symmetry, as can be seen from the previous section, is an
essential ingredient in the phenomenological analysis of the $KN$ system,
apart from obvious mass--difference effects, apparent only in the close
proximity of the thresholds, which one can describe by modifying the
K--matrix formalism as outlined above$^{10}$.

One way to check isospin symmetry is to relate the amplitudes derived from
{\sl charged} kaon scattering to the data from $K^0_S$
regeneration. Since isospin relates the scattering of charged kaons on protons
to the regeneration on neutrons (and {\sl vice versa}), the test is better
performed on an isoscalar nuclear target, such as deuterium or $^4$He. We
should have indeed, apart from kinematical corrections and CP--violation
effects,
$$ T(K_L^0p \to K_S^0p) = {1\over2} [T(K^0p \to K^0p) - T(\overline{K}^0p \to
\overline{K}^0p)] =$$
$$= {1\over2} [T(K^+n \to K^+n) - T(K^-n \to K^-n)] \eqno(27)$$
and
$$ T(K_L^0n \to K_S^0n) = {1\over2} [T(K^0n \to K^0n) - T(\overline{K}^0n \to
\overline{K}^0n)] =$$
$$= {1\over2} [T(K^+p \to K^+p) - T(K^-p \to K^-p)] \ ;\eqno(28)$$
when we introduce these equalities in a nuclear scattering calculation, as in
{\sl e.g.} a Glauber model, all {\sl elastic} multiple scattering effects
should apply equally to both the right-- and left--hand sides of the
equalities for an isoscalar nucleus, protecting the identity from a large
fraction of the ``nuclear'' effects$^{38}$.

Up to now such tests would have been possible only at higher momenta, where
the opening of inelastic channels in the $S=+1$ systems complicates
calculations further: a test performed in the elastic region of this system
should make things simpler and clearer, at least in the $S=+1$ sector.

The second problem, related in many theoretical analyses to observations from
inelastic electron and muon scattering on nuclei, namely to changes in the
electromagnetic properties and in the deep--inelastic structure functions of
nucleons bound in nuclei with respect to the free ones, is the
``antishadowing'' effect observed at momenta around $800\ MeV/c$ for $K^+$
scattering on nuclear targets$^{39}$.
Conventional Glauber--model calculations$^{40}$ led to
expect a ratio $(2\sigma_A)/(A\sigma_D)$ slightly less than unity and
decreasing with both the kaon momentum and the target mass number $A$,
while the measured
values were larger than unity and increasing with momentum.
This led to think,
as an explanation of this and of the aforementioned
electromagnetic phenomena, of
a ``swelling'' of the bound nucleons with respect to free ones, in line with
some of the explanations put forward for the ``nuclear'' EMC effect, though at
a much higher energy scale$^{41}$.

New data have recently confirmed this trend$^{42}$,
but only for momenta higher than approximately $600\ MeV/c$; a
possibility coming to mind is that the opening of inelastic channels, such as
$\pi KN$ (or more simply quasi--two--body ones as $KN^*$ and $K^*N$),
might necessitate the
introduction of inelastic intermediate states absent in a conventional
Glauber--model calculation, phenomenon analogous to the need to introduce
inelastic diffraction in the intermediate steps of a multiple--scattering
formalism to explain diffractive
processes on nuclei at much higher energies:
thus the data would be just showing the opening of the
threshold for such a phenomenon, particularly visible in the $K^+$--scattering
case because of the extremely long mean--free path of this hadron in nuclear
matter (about $7\ fm$).

Measurements of the $K^+$ cross sections on different nuclei in DA$\Phi$NE's
kinematical region, where $K^+N$ interactions are purely elastic, should help
close the issue when compared with accurate Glauber--model
calculations$^{40}$.

We would like to close reminding the reader that information on the $S=+1,\
I=0$ channel in this energy region is coming {\sl entirely} from
extrapolations from higher--momen-tum data, since $K^+$--scattering (and
regeneration) data on deuterium are {\sl not} available at momenta lower than
about $300\ MeV/c$: at present we have only a generic idea about the order of
magnitude of the {\sl absolute value} of the $KN$ $I=0$ scattering length,
expected to be of the order of some times $10^{-2}\ fm$ from forward
dispersion relations and the lowest--momentum regeneration data$^{33,34}$.
An accurate measurement of the cross sections for $K^+$ {\sl incoherent}
scattering on deuterium, possible at
DA$\Phi$NE over a wide angular range, would thus give us the first {\sl
direct} measurement of this quantity.

\vglue 2.0truecm

\centerline{\bf REFERENCES AND FOOTNOTES }

\vglue 0.3truecm

\item{1.} $K^\pm p$ data: W.E. Humphrey and R.R. Ross: {\sl Phys. Rev.}
{\bf 127} (1962) 1; G.S. Abrams and B. Sechi--Zorn: {\sl Phys. Rev.} {\bf 139}
(1965) B 454; M. Sakitt, {\sl et al.: Phys Rev} {\bf 139} (1965) B 719; J.K.
Kim: Columbia Univ. report {\sl NEVIS--149} (1966), and {\sl Phys. Rev. Lett.}
{\bf 14} (1970) 615; W. Kittel, G Otter and I. Wa\v{c}ek: {\sl Phys. Lett}
{\bf 21} (1966) 349; D. Tovee, {\sl et al.: Nucl. Phys.} {\bf B 33} (1971)
493; T.S. Mast, {\sl et al.: Phys. Rev.} {\bf D 11} (1975) 3078, and {\bf D
14} (1976) 13; R.O. Bargenter, {\sl et al.: Phys. Rev.} {\bf D 23} (1981)
1484. $K^-d$ data: R. Armenteros, {\sl et al.: Nucl. Phys.} {\bf B 18} (1970)
425.

\item{2.} TST Collaboration: R.J. Novak, {\sl et al.: Nucl Phys.} {\bf B 139}
(1978) 61; N.H. Bedford, {\sl et al.: Nukleonika} {\bf 25} (1980) 509; M.
Goossens, G. Wilquet, J.L. Armstrong and J.H. Bartley: {\sl ``Low and
Intermediate Energy Kaon--Nucleon Physics''}, ed. by E. Ferrari and G. Violini
(D. Reidel, Dordrecht 1981), p. 131; J. Ciborowski, {\sl et al.: J. Phys.}
{\bf G 8} (1982) 13; D. Evans, {\sl et al.: J. Phys.} {\bf G 9} (1983) 885; J.
Conboy, {\sl et al.: J. Phys} {\bf G 12} (1986) 1143. A good description of
the experiment is in D.J. Miller, R.J. Novak and T. Tyminiecka: {\sl ``Low and
Intermediate Energy Kaon-Nucleon Physics''}, ed by E. Ferrari and G. Violini
(D. Reidel, Dordrecht 1981), p. 251.

\item{3.} $K^0_Lp$ data: J.A. Kadyk, {\sl et al.: Phys. Lett.} {\bf 17} (1966)
599, and report {\sl UCRL--18325} (1968); R.A. Donald, {\sl et al.: Phys.
Lett.} {\bf 22} (1966) 711; G.A. Sayer, {\sl et al.: Phys. Rev.} {\bf 169}
(1968) 1045.

\item{4.} R.H. Dalitz and A. Deloff: {\sl J. Phys.} {\bf G 17} (1991) 289,
erratum {\bf G 19} (1993) 1423. See also ref. 24 for a wider bibliography
on this subject.

\item{5.} L. Montanet, {\sl et al.} (Particle Data Group): {\sl Phys. Rev.}
{\bf D 50} (1994) 1173.

\item{6.} For conventions and kinematical notations we have adopted the same
as: G. H\"ohler, F. Kaiser, R. Koch and E. Pietarinen: {\sl ``Handbook of
Pion--Nucleon Scattering''} (Fachinformationszentrum, Karlsruhe 1979), and
{\sl ``Landolt--B\"ornstein, New Series, Group I, Vol. 9b''}, ed. by H.
Schopper (Springer--Verlag, Berlin 1983), which have become a ``standard''
for describing $\pi N$ scattering.

\item{7.} J. Hamiltom, I. \O verb\o\ and B. Tromborg: {\sl Nucl. Phys.} {\bf B
60} (1973) 443; B. Tromborg and J. Hamilton: {\sl Nucl. Phys.} {\bf B 76}
(1974) 483; J. Hamilton: {\sl Fortschr. Phys.} {\bf 23} (1975) 211.

\item{8.} B. Tromborg, S. Waldenstr\o m and I. \O verb\o : {\sl Ann. Phys.
(N.Y.)} {\bf 100} (1976) 1; {\sl Phys. Rev.} {\bf D 15} (1977) 725; {\sl Helv.
Phys. Acta} {\bf 51} (1978) 584.

\item{9.} G.C. Oades and G. Rasche: {\sl Helv. Phys. Acta} {\bf 44} (1971) 5,
and {\sl Phys. Rev.} {\bf D 4} (1971) 2153; G. Rasche and W.S. Woolcock: {\sl
Helv. Phys. Acta} {\bf 45} (1972) 642; H. Zimmermann: {\sl Helv. Phys. Acta}
{\bf 45} (1973) 1117, {\bf 47} (1974) 30, and {\bf 48} (1975) 191.

\item{10.} G. Rasche and W.S. Woolcock: {\sl Helv. Phys. Acta} {\bf 49} (1976)
435, 455, 557, and {\bf 50} (1977) 407; {\sl Fortschr. Phys.} {\bf 25} (1977)
501.

\item{111.} R. Seki: {\sl ``Meson-Nuclear Physics 1976''}, ed. by P.D. Barnes,
R.A. Eisenstein and L.S. Kisslinger (A.I.P., New York 1976), p. 80; M.
Schechter: {\sl Ann. Phys. (N.Y.)} {\bf 101} (1976) 601.

\item{12.} P.M. Gensini: {\sl Lett. Nuovo Cimento} {\bf 38} (1983) 620; {\sl
Nuovo Cimento} {\bf A 78} (1983) 471; P.M. Gensini and G.R. Semeraro: {\sl
``Perspectives on Theoretical Nuclear Physics (I)''}, ed. by L. Bracci, {\sl
et al.} (E.T.S. Ed., Pisa 1986), p 91.

\item{13.} M. Atarashi, K. Hira and H. Narumi: {\sl Prog. Theor. Phys.} {\bf
60} (1978) 209; J.R. Rook: {\sl Nucl. Phys.} {\bf A 326} (1979) 244.

\item{14.} J.D. Davies, {\sl et al.}: {\sl Phys. Lett.} {\bf B 83} (1979) 55.

\item{15.} M. Izycki, {\sl et al.}: {\sl Z. Phys.} {\bf A 297} (1980) 11; P.M.
Bird, {\sl et al}: {\sl Nucl. Phys.} {\bf A 404} (1983) 482.

\item{16.} A. Deloff and J. Law: {\sl Phys. Rev.} {\bf C 20} (1979) 1597; R.C.
Barrett: {\sl J. Phys.} {\bf G 8} (1982) L 39, erratum {\bf G 9} (1983) 355.
See also the reviews presented at the Legnaro Hypernuclear Conference by C.J.
Batty and A. Gal: {\sl Nuovo Cimento} {\bf A 102} (1989) 255, and at the
Cambridge, MA., PANIC by C.J. Batty: {\sl Nucl. Phys.} {\bf A 508} (1990) 89c.

\item{17.} Some of the ``unreasonable explanations'' include: K.S. Kumar and
Y. Nogami: {\sl Phys. Rev.} {\bf D 21} (1980) 1834; K.S. Kumar, Y. Nogami, W.
van Dijk and D. Kiang: {\sl Z. Phys.} {\bf A 304} (1982) 301; D. Kiang, K.S.
Kumar, Y. Nogami and W. van Dijk: {\sl Phys. Rev.} {\bf C 30} (1984) 1638; K.
Tanaka and A. Suzuki: {\sl Phys. Rev.} {\bf C 45} (1992) 2068.

\item{18.} G. Reifenr\"other and E. Klempt: {\sl Phys. Lett.} {\bf B 248}
(1990) 250.

\item{19.} M. Iwasaki: {\sl ``Workshop on Science at the KAON Factory''}, ed.
by D.R. Gill (TRIUMF, Vancouver 1991), Vol. 2, p. 145.

\item{20.} J.L. Friar: {\sl Z. Phys.} {\bf A 292} (1979) 1, erratum {\bf A
303} (1981) 84; {\bf A 297} (1980) 147. For the general formalism the best
description is still to be found in: H. Feshbach and F. Villars: {\sl
Rev. Mod. Phys.} {\bf 30} (1958) 24.

\item{21.} S. Deser, M.L. Goldberger, K. Baumann and W. Thirring: {\sl Phys.
Rev.} {\bf 96} (1954) 774. The formula has been re--derived under much general
assumptions, and verified extremely well for pionic hydrogen. See: J. Thaler
and H.F.K. Zingl: {\sl J. Phys.} {\bf G 8} (1982) 771; J. Thaler: {\sl J.
Phys.} {\bf G 9} (1983) 1009; W.B. Kaufmann and W.R. Gibbs: {\sl Phys. Rev.}
{\bf C 35} (1987) 838.

\item{22.} R.J. Hemingway: {\sl Nucl. Phys.} {\bf B 253} (1985) 742. Older
data are even poorer in statistics: see ref. 31 for a comparison. See also,
for formation on bound nucleons, B. Riley, I.T. Wang, J.G. Fetkovich and J.M.
McKenzie: {\sl Phys. Rev.} {\bf D 11} (1975) 3065.

\item{23.} G.C. Oades: {\sl Nuovo Cimento} {\bf 102 A} (1989) 237.

\item{24.} P.M. Gensini and G. Soliani: {\sl Lett. Nuovo Cimento} {\bf 4}
(1970) 329; A.D. Martin, B.R. Martin and G.G. Ross: {\sl Phys. Lett.} {\bf B
35} (1971) 62; P.N. Dobson jr. and R. McElhaney: {\sl Phys. Rev.} {\bf D 6}
(1972) 3256; G.C. Oades and G. Rasche: {\sl Nuovo Cimento} {\bf 42 A} (1977)
462; R.H. Dalitz and J.G. McGinley: {\sl ``Low and Intermediate Energy
Kaon--Nucleon Physics''}, ed. by E. Ferrari and G. Violini (D. Reidel,
Dordrecht 1981), p. 381, and the Ph.D. thesis by McGinley (Oxford Univ. 1979);
G.C. Oades and G. Rasche: {\sl Phys. Scr.} {\bf 26} (1982) 15; J.P. Liu:
{\sl Z. Phys.} {\bf C 22} (1984) 171; B.K. Jennings: {\sl Phys. Lett.}
{\bf B 176} (1986) 229; P.B. Siegel and W. Weise: {\sl Phys. Rev.} {\bf C 38}
(1988) 2221. Even more recent theoretical models are to be found in:
M. Arima, S. Matsui and K. Shimizu: {\sl Phys. Rev.} {\bf C 39} (1994) 2831;
C.L. Schat, N.N. Scoccola and C. Gobbi: Univ. Pavia report {\sl
TAN--FNT--94--09}, also available as hep--ph/9408360.

\item{25.} B.L. Roberts: {\sl Nucl Phys} {\bf A 479} (1988) 75c; B.L. Roberts,
{\sl et al.: Nuovo Cimento} {\bf 102 A} (1989) 145; D.A. Whitehouse, {\sl et
al.: Phys. Rev. Lett.} {\bf 63} (1989) 1352.

\item{26.} See the review presented at the Legnaro Hypernuclear Conference by
J. Lowe: {\sl Nuovo Cimento} {\bf 102 A} (1989) 167.

\item{27.} J.W. Darewich, R. Koniuk and N. Isgur: {\sl Phys. Rev.} {\bf D 32}
(1985) 1765; H. Burkhardt, J. Lowe and A.S. Rosenthal: {\sl Nucl Phys.} {\bf A
440} (1985) 653; R.L. Workman and H.W. Fearing: {\sl Phys. Rev.} {\bf D 37}
(1988) 3117; R.A. Williams, C.R. Ji and S. Cotanch: {\sl Phys. Rev.} {\bf D
41} (1990) 1449; {\sl Phys. Rev.} {\bf C 43} (1991) 452; H. Burkhardt and J.
Lowe: {\sl Phys. Rev.} {\bf C 44} (1991) 607. For radiative capture on
deuterons (and other light nuclei), see: R.L. Workman and H.W. Fearing: {\sl
Phys. Rev.} {\bf C 41} (1990) 1688; C. Bennhold: {\sl Phys. Rev.} {\bf C 42}
(1990) 775.

\item{28.} Y.S. Zhong, A.W. Thomas, B.K. Jennings and R.C. Barrett: {\sl Phys.
Lett.} {\bf B 171} (1986) 471; {\sl Phys. Rev.} {\bf D 38} (1988) 837 (which
corrects a numerical error contained in the previous paper).

\item{29.} A more recent analyis including accurate consideration of the
initial--state interaction has been performed by P.B. Siegel and B. Saghai:
Saclay report {\sl DAPNIA--SphN--94--64}, subm to {\sl Phys. Rev.} {\bf C}.

\item{30.} See the review presented by B.R. Martin at the 1972 Ba\v{s}ko Polje
International School, published in: {\sl ``Textbook on Elementary Particle
Physics. Vol. 5: Strong Interactions''}, ed by M. Nikoli\v{c} (Gordon and
Breach, Paris 1975).

\item{31.} R.H. Dalitz and S.F. Tuan: {\sl Ann. Phys. (N.Y.)} {\bf 10} (1960)
307.

\item{32.} Y.A. Chao, R. Kr\"amer, D.W. Thomas and B.R. Martin: {\sl Nucl.
Phys.} {\bf B 56} (1973) 46.

\item{33.} A.D. Martin: {\sl Phys. Lett.} {\bf B 65} (1976) 346.

\item{34.} A.D. Martin: {\sl ``Low and Intermediate Kaon--Nucleon Physics''},
ed. by E. Ferrari and G. Violini (D. Reidel, Dordrecht 1981), p. 97; {\sl
Nucl. Phys.} {\bf B 179} (1981) 33.

\item{35.} R.H. Dalitz, J. McGinley, C. Belyea and S. Anthony: {\sl
``Proceedings of the International Conference on Hypernuclear and Kaon
Physics''}, ed. by B. Povh (M.P.I., Heidelberg 1982), p. 201.

\item{36.} G.K. Atkin, B. Di Claudio, G. Violini and N.M. Queen: {\sl Phys.
Lett.} {\bf B 95} (1980) 447; {\sl ``Low and Intermediate Energy Kaon--Nucleon
Physics''}, ed. by E. Ferrari and G. Violini (D. Reidel, Dordrecht 1981), p.
131; J. Antol\'\i{n}: {\sl Phys. Rev.} {\bf D 43} (1991) 1532.

\item{37.} See the discussion on this point by D.J. Miller in: {\sl
``Proceedings of the International Conference on Hypernuclear and Kaon
Physics''}, ed. by B. Povh (M.P.I., Heidelberg 1982), p. 215.

\item{38.} See the talk by V.L. Telegdi, in: {\sl ``High--Energy Physics and
Nuclear Structure''}, ed. by D.E. Nagle, {\sl et al.} (A.I.P., New York 1975),
p. 289.

\item{39.} E. Piasetzsky: {\sl Nuovo Cimento} {\bf 102 A} (1989) 281; Y.
Mardor, {\sl et al.: Phys. Rev. Lett.} {\bf 65} (1990) 2110. Older data at
higher momenta are to be found in: D.V. Bugg, {\sl et al.: Phys. Rev.} {\bf
168} (1968) 1466.

\item{40.} P.B. Siegel, W.B. Kaufmann and W.R. Gibbs: {\sl Phys. Rev.} {\bf C
30} (1984) 1256; Ya.A. Berdnikov, A.M. Makhov and V.I. Ostroumov: {\sl Sov. J.
Nucl. Phys.} {\bf 49} (1989) 618; Ya.A. Berdnikov and A.M. Makhov: {\sl Sov.
J. Nucl. Phys.} {\bf 51} (1990) 579.

\item{41.} P.B. Siegel, W.B. Kaufmann and W.R. Gibbs: {\sl Phys. Rev.} {\bf C
31} (1985) 2184; G.E. Brown, C.B. Dover, P.B. Siegel and W. Weise: {\sl Phys.
Rev. Lett.} {\bf 60} (1988) 2723; W.B. Kaufmann and W.R. Gibbs: {\sl Phys.
Rev.} {\bf C 40} (1989) 1729; W. Weise: {\sl Nuovo Cimento} {\bf 102 A} (1989)
265; J. Labarsouque: {\sl Nucl. Phys.} {\bf A 506} (1990) 539; M. Mizoguchi
and H. Toki: {\sl Nucl. Phys.} {\bf A 513} (1990) 685; J.C. Caillon and J.
Labarsouque: {\sl Phys. Lett.} {\bf B 295} (1992) 21; {\sl Phys. Rev.} {\bf C
45} (1992) 2503; {\sl J. Phys.} {\bf G 19} (1993) L 117; {\sl Phys. Lett.}
{\bf B 311} (1993) 19; {\sl Nucl. Phys.} {\bf A 572} (1994) 649, erratum {\bf
A 576} (1994) 639.

\item{42.} J. Alster, {\sl et al.}: {\sl Nucl. Phys.} {\bf A 547} (1992) 321c.

\bye